\documentclass[sigconf]{acmart}
\AtBeginDocument{%
  }

\setcopyright{acmlicensed}
\copyrightyear{2018}
\acmYear{2018}
\acmDOI{XXXXXXX.XXXXXXX}
\acmConference[Conference acronym 'XX]{Make sure to enter the correct
  conference title from your rights confirmation email}{June 03--05,
  2018}{Woodstock, NY}
\acmISBN{978-1-4503-XXXX-X/2018/06}

\usepackage{tikz}
\usetikzlibrary{arrows.meta, positioning}

\begin{document}

\title[OpenProposal]{OpenProposal \\ Platform for Transparent Research Funding Review}

\author{Sakshi Ahuja}
\email{sakshi.ahuja@niser.ac.in}
\affiliation{%
  \institution{DST Center for Policy Research \\ National Institute of Science Education and Research (NISER)}
  \city{Bhubaneswar}
  \state{Odisha}
  \country{India}
}

\author{Subhankar Mishra}
\email{smishra@niser.com}
\affiliation{%
  \institution{DST Center for Policy Research \\ National Institute of Science Education and Research (NISER)}
  \city{Bhubaneswar}
  \state{Odisha}
  \country{India}
}

\renewcommand{\shortauthors}{Ahuja and Mishra}

\begin{abstract}
Research funding allocation remains a critical bottleneck in scientific advancement, yet the review process for funding proposals lacks the transparency that has revolutionized academic paper peer review. Traditional funding agencies operate with closed review systems, limiting accountability and preventing systematic improvements. We present OpenProposal, a proof-of-concept web-based platform that explores how transparency principles from OpenReview might be adapted to research funding proposal evaluation. Built using modern web technologies including Next.js , React , and Prisma , OpenProposal demonstrates the technical feasibility of public reviews, author rebuttals, and transparent decision-making while attempting to protect sensitive information such as budgets. Our platform prototype addresses key limitations identified in current funding systems by providing mechanisms for community engagement, reviewer accountability, and potential data-driven insights into peer review processes. Through system design and implementation, we explore how transparent funding review could potentially enhance scientific integrity and improve research funding decisions, though empirical validation remains necessary. This work contributes a technical foundation for transparent funding review and identifies design considerations for future research on peer review mechanisms in funding contexts. OpenProposal is available for testing at \url{https://smlab.niser.ac.in/project/openproposal/}. Code is available at \url{https://github.com/smlab-niser/openproposal       } 
\end{abstract}



\keywords{peer review, research funding, transparency, web platform, scientific integrity, machine learning, open science, proposal evaluation}


\maketitle
\section{Introduction}

Peer review serves as the cornerstone of scientific quality assurance, governing not only the publication of research papers but also the allocation of research funding. In the funding domain, reviewers evaluate proposals to identify the most promising research based on novelty, feasibility, and potential impact. However, unlike the recent transparency trends in academic publishing, the review process for research funding remains largely opaque, with reviewers maintaining anonymity and decisions made behind closed doors.

This lack of transparency in funding review systems presents several critical challenges. Principal Investigators (PIs) receive limited feedback on rejected proposals, hindering their ability to improve future submissions. The absence of public accountability may lead to inconsistent review quality and potential bias. Furthermore, the research community loses valuable insights that could emerge from analyzing funding decision patterns and reviewer feedback.

Major funding agencies worldwide, including the Department of Science and Technology (DST) in India and the National Science Foundation (NSF) in the United States, operate predominantly closed review systems. While these agencies have served the scientific community for decades, their opaque processes limit learning opportunities and may inadvertently perpetuate systematic biases in funding allocation.

The success of OpenReview in transforming academic conference peer review demonstrates the viability of transparent review systems. OpenReview's adoption by premier machine learning venues such as ICLR and NeurIPS has shown that public reviews, author rebuttals, and open discussions can enhance review quality while maintaining scientific rigor. This success motivates our investigation into applying similar transparency principles to research funding.

We present OpenProposal, a web-based platform that extends OpenReview's transparency model to research funding proposal evaluation. Our system enables public reviews, author rebuttals, and community engagement while preserving the confidentiality of sensitive information such as budget details. Built using modern web technologies, OpenProposal provides a scalable solution for transparent funding review that can complement existing agency workflows.

Our contributions include: 
\begin{enumerate}
    \item A proof-of-concept web-based platform \textbf{OpenProposal} that demonstrates how OpenReview's transparency principles can be adapted to research funding proposal evaluation, addressing the opacity limitations of current funding review systems
    \item Design mechanisms for selectively preserving confidentiality of sensitive funding information (such as budgets) while enabling public review of research merit, author rebuttals, and community engagement
    \item A technical foundation that enables future empirical analysis of funding review patterns, reviewer feedback quality, and decision-making processes that are currently inaccessible in closed review systems
\end{enumerate}

\begin{figure*}[htbp]
    \centering
    \includegraphics[width=\linewidth]{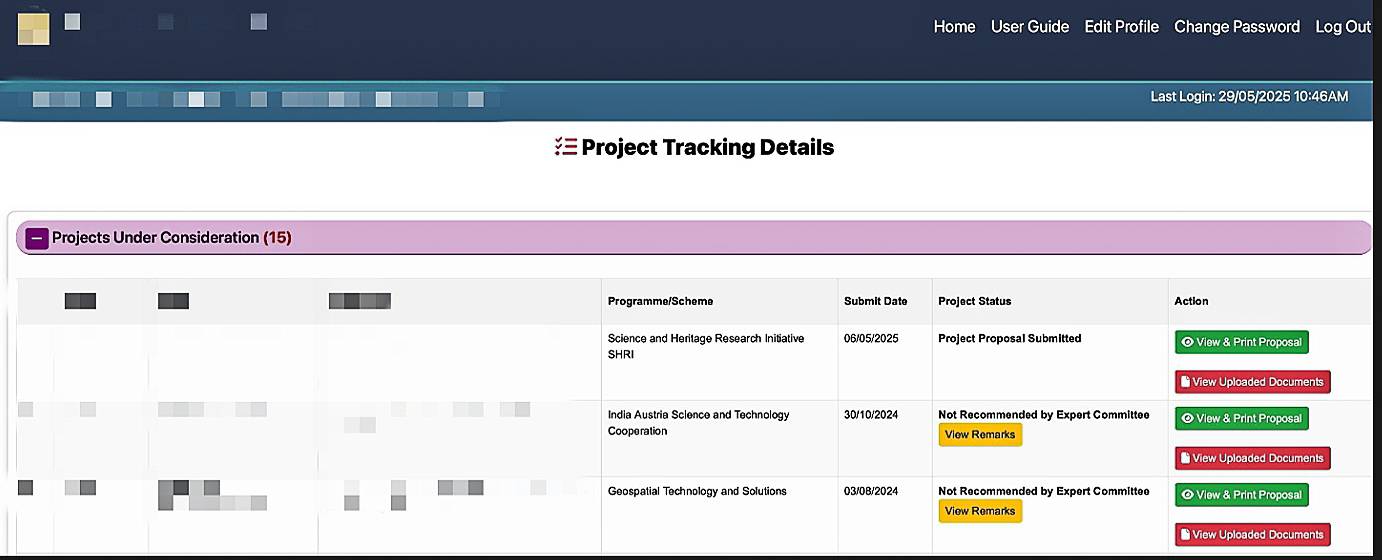}
    \caption{Dashboard view of project proposals under consideration in the DST portal, showing submission details, review status, and access to proposal documents.}
    \label{fig:dstdash}
\end{figure*}

\begin{figure}[htbp]
    \centering
    \includegraphics[width=\linewidth]{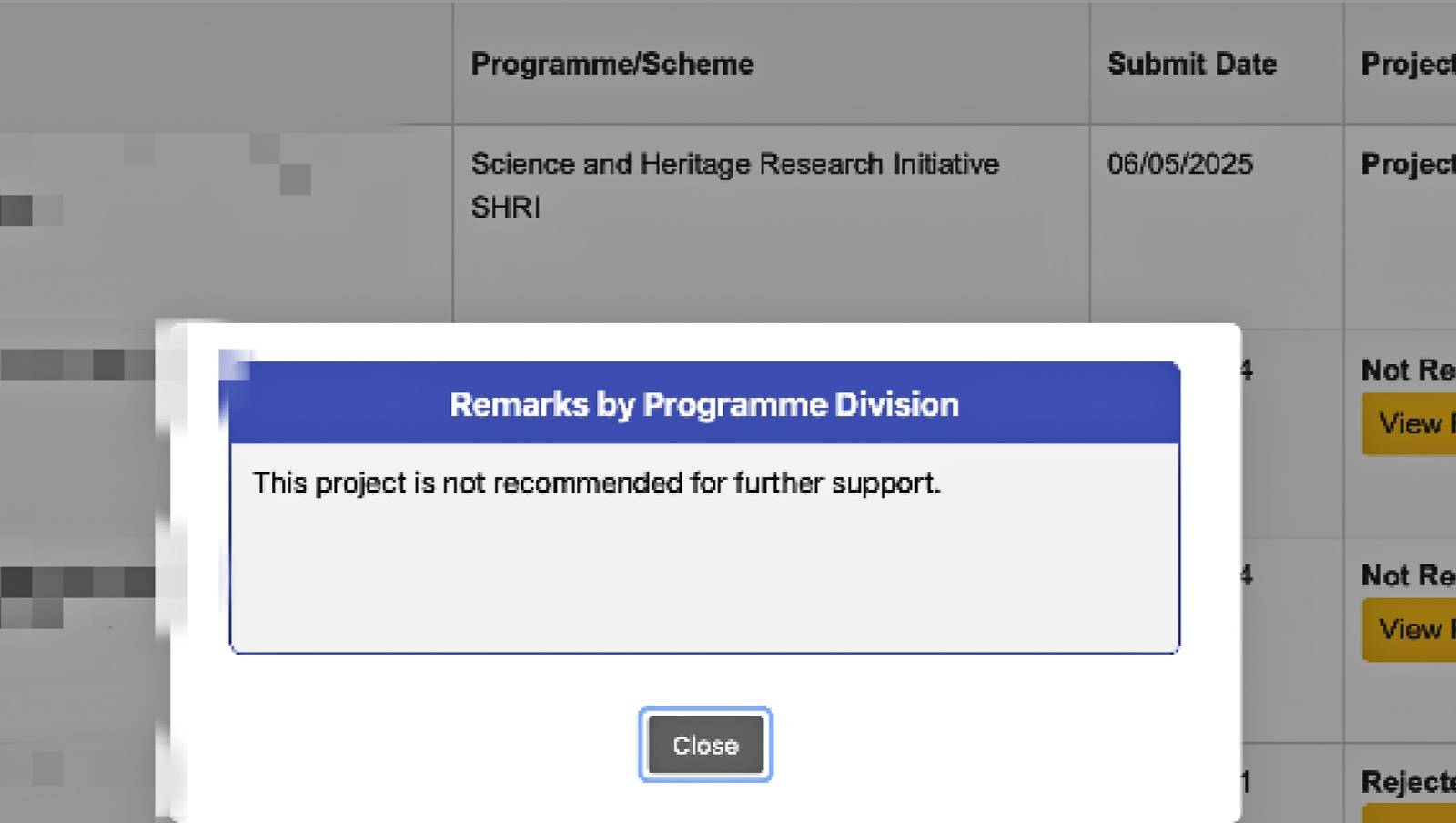}
    \caption{Screenshot of the Department of Science and Technology (DST) project portal displaying proposal status and programme division remarks but the remarks are too short and uninformative that doesn't help the PI's to improve proposals.}
    \label{fig:dstremarks}
\end{figure}

\section{Related Work}

\subsection{Transparent Peer Review Systems}

The OpenReview platform~\cite{openreview2013} revolutionized academic peer review by introducing transparency to conference review processes. Adopted by leading machine learning conferences including ICLR \cite{iclr2025} and NeurIPS \cite{neurips2025}, OpenReview enables public access to submissions, reviews, author rebuttals, and final decisions \cite{wang2023have}. This transparency has been shown to improve review quality, reduce bias, and facilitate community engagement in the review process~\cite{shah2018design}.

Studies on OpenReview's impact demonstrate several benefits of transparent review. Reviewers tend to provide more constructive and detailed feedback when their reviews are public~\cite{tomkins2017reviewer}. The ability for authors to provide rebuttals leads to more informed decision-making, particularly for borderline papers~\cite{lee2013bias}. Additionally, the availability of review data has enabled meta-research on peer review quality and bias patterns~\cite{kang2018dataset}. 

While OpenReview has successfully transformed the way academic papers are submitted and reviewed in conferences by making the process more transparent and interactive, it is not designed for managing research funding proposals. OpenReview focuses on academic publishing: authors submit papers, reviewers give feedback, and a final decision is made for conference publication. In contrast, funding proposals involve a more administrative and policy-driven workflow—managing budgets, eligibility checks, agency-level approvals, and longer-term project outcomes.

While the front-end of OpenReview is open source, we encountered difficulties in reproducing its full functionality, as the backend skeleton setup is not publicly accessible, making it hard to customize or extend the platform for funding workflows. This limitation further motivated the development of OpenProposal, which is tailored specifically for research grant calls and built from the ground up to support structured funding workflows.

\subsection{Funding Agency Review Systems}

Traditional funding agencies operate with closed review processes that prioritize reviewer anonymity and confidential decision-making. The National Science Foundation (NSF) \cite{nsf2025} employs a merit review system based on intellectual merit and broader impacts, but reviewer identities and detailed feedback remain confidential~\cite{nsf2020merit}. Similarly, the Department of Science and Technology (DST) \cite{dst2025} through onlinedst portal as shown in Fig \ref{fig:dstdash} in India follows a closed review process where proposals are evaluated by expert panels, but the review process lacks transparency as shown in Figure \ref{fig:dstremarks} ~\cite{dst2021guidelines}.

International funding agencies such as the European Research Council (ERC)\cite{erc2025}, UK Research and Innovation (UKRI) \cite{ukri2025}, and the Japan Society for the Promotion of Science (JSPS) \cite{jsps2025} follow similar closed review models. While these systems have supported significant scientific advances, they provide limited opportunities for applicants to learn from feedback or for the community to understand funding allocation patterns.

\subsection{Challenges in Funding Review}

Research on funding review processes has identified several systemic issues. The lack of feedback to unsuccessful applicants limits their ability to improve future proposals~\cite{guthrie2018grant}. Additionally, the absence of public data on funding decisions prevents systematic analysis of review quality and fairness~\cite{wessely1998peer}.

Some agencies have attempted to address these issues through incremental reforms. The NSF now provides summary statements to declined proposals, and some agencies have experimented with modified review formats. However, these changes fall short of the transparency achieved in academic publishing through platforms like OpenReview.

\section{System Design and Implementation}

\begin{figure}[h]
    \centering
    \includegraphics[width=0.4\linewidth]{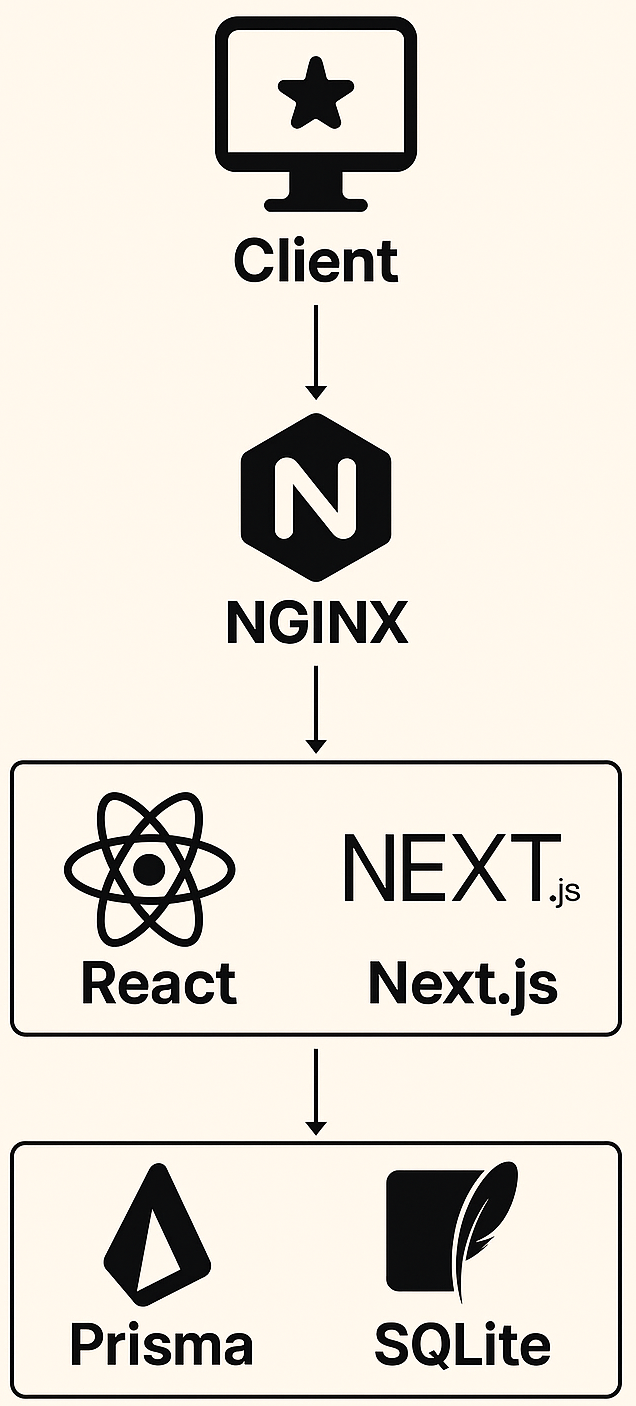 }
    \caption{System architecture of the OpenProposal platform. The client interacts via a React-based frontend served by NGINX. Next.js handles both frontend and backend logic, interfacing with a Prisma ORM connected to an SQLite database for data persistence..}
    \label{fig:architecture}
\end{figure}

\subsection{System Architecture}
OpenProposal employs a modern full-stack web architecture built on Next.js \cite{nextjs2025} and React \cite{react2025} to deliver a responsive and scalable platform for transparent funding review Figure~\ref{fig:architecture}. The system architecture follows a three-tier design pattern comprising a presentation layer (React-based user interface), application layer (Next.js API routes), and data persistence layer (Prisma ORM with PostgreSQL database).

The frontend utilizes React's component-based architecture to create reusable UI elements for proposal submission, review interfaces, and result visualization. Next.js provides server-side rendering capabilities, improving performance and search engine optimization while enabling seamless navigation between platform sections. The framework's API routes facilitate secure communication between the client and server, handling authentication, authorization, and data validation.

For data management, we employ Prisma \cite{prisma2025} as our Object-Relational Mapping (ORM) toolkit, which provides type-safe database access and automated migrations. The database schema supports complex relationships between users, proposals, reviews, funding calls, and institutional affiliations while maintaining referential integrity and supporting efficient queries for large scale data analysis.

The platform implements role-based access control to manage different user types (Principal Investigators, Reviewers, Agency Representatives, Community Members) with appropriate permissions for each workflow stage. Security measures include JWT-based authentication \cite{jwt2025}, input sanitization, and data encryption for sensitive information.

\subsection{Workflow Design}
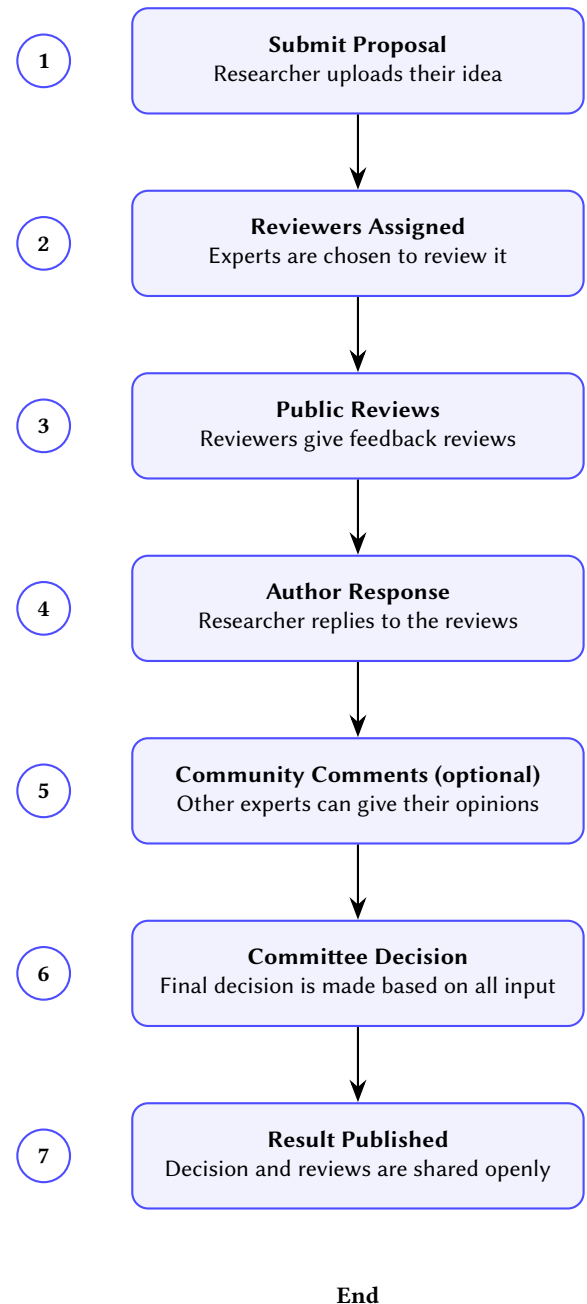
\begin{figure}[htbp]
  \centering
  \begin{tikzpicture}[
      node distance=1cm,
      every node/.style={font=\sffamily},
      box/.style={
          draw=blue!70,
          thick,
          rounded corners=6pt,
          minimum width=6cm,
          minimum height=1.4cm,
          fill=blue!5,
          align=center
      },
      num/.style={
          draw=blue!70,
          circle,
          thick,
          fill=white,
          minimum size=7mm,
          font=\bfseries
      },
      arrow/.style={
          -{Stealth[length=3mm]},
          thick
      }
  ]

  \node[font=\bfseries\Large] at (0,2.2) {OpenProposal Flowchart};

  \node[box] (s1) {
      \textbf{Submit Proposal}\\
      Researcher uploads their idea
  };
  \node[num, left=8mm of s1.west] {1};

  \node[box, below=of s1] (s2) {
      \textbf{Reviewers Assigned}\\
      Experts are chosen to review it
  };
  \node[num, left=8mm of s2.west] {2};

  \node[box, below=of s2] (s3) {
      \textbf{Public Reviews}\\
      Reviewers give feedback reviews
  };
  \node[num, left=8mm of s3.west] {3};

  \node[box, below=of s3] (s4) {
      \textbf{Author Response}\\
      Researcher replies to the reviews
  };
  \node[num, left=8mm of s4.west] {4};

  \node[box, below=of s4] (s5) {
      \textbf{Community Comments (optional)}\\
      Other experts can give their opinions
  };
  \node[num, left=8mm of s5.west] {5};

  \node[box, below=of s5] (s6) {
      \textbf{Committee Decision}\\
      Final decision is made based on all input
  };
  \node[num, left=8mm of s6.west] {6};

  \node[box, below=of s6] (s7) {
      \textbf{Result Published}\\
      Decision and reviews are shared openly
  };
  \node[num, left=8mm of s7.west] {7};

  \node[below=0.9cm of s7, font=\bfseries] {End};

  \draw[arrow] (s1) -- (s2);
  \draw[arrow] (s2) -- (s3);
  \draw[arrow] (s3) -- (s4);
  \draw[arrow] (s4) -- (s5);
  \draw[arrow] (s5) -- (s6);
  \draw[arrow] (s6) -- (s7);

  \end{tikzpicture}
  \caption{OpenProposal workflow showing the complete process from proposal submission to transparent result publication.}
  \label{fig:workflow}
\end{figure}

Our workflow design tries to strike a careful balance between the transparency we want to achieve and the practical realities of funding review. Figure~\ref{fig:workflow} shows how the process flows from initial submission through to the final publication of results.

When researchers submit proposals, they fill out structured forms covering the usual elements: research goals, methods, timeline, and budget details (when agencies allow this information to be shared). We automatically check that submissions are complete and properly formatted before passing them along for review.

For reviewer assignment, we match experts to proposals using algorithms that look at research areas, keywords, and how well reviewers have performed in the past. Some agencies prefer to manage reviewer assignments themselves, while others are open to community suggestions for who might be good reviewers. Either way, each proposal gets multiple independent reviews using consistent criteria, though we allow some flexibility for different research domains.

One thing we're particularly excited about is letting authors respond to reviews. This addresses a major frustration many researchers have with current funding systems - when reviewers misunderstand something or make an error, there's usually no way to correct it. In our system, Principal Investigators can clarify their methods, correct factual mistakes, or provide additional evidence. This back-and-forth tends to be especially helpful for proposals that are right on the borderline for funding.

The final decision process brings together all the reviews, author responses, and any additional community input. Program officers can see the full discussion history and how the proposal evolved through the review process. Once decisions are made, we publish the outcomes along with the review discussions (anonymized where needed) so the whole community can learn from the process.


\subsection{User Interface Design}

We designed the OpenProposal interface with academic users in mind, trying to make complex review workflows as straightforward as possible. Our approach builds on standard web usability principles but adapts them for the specific needs of researchers, reviewers, and funding agencies.

\subsubsection{Authentication and User Management}
\begin{figure}[h]
    \centering
    \includegraphics[width=0.8\linewidth]{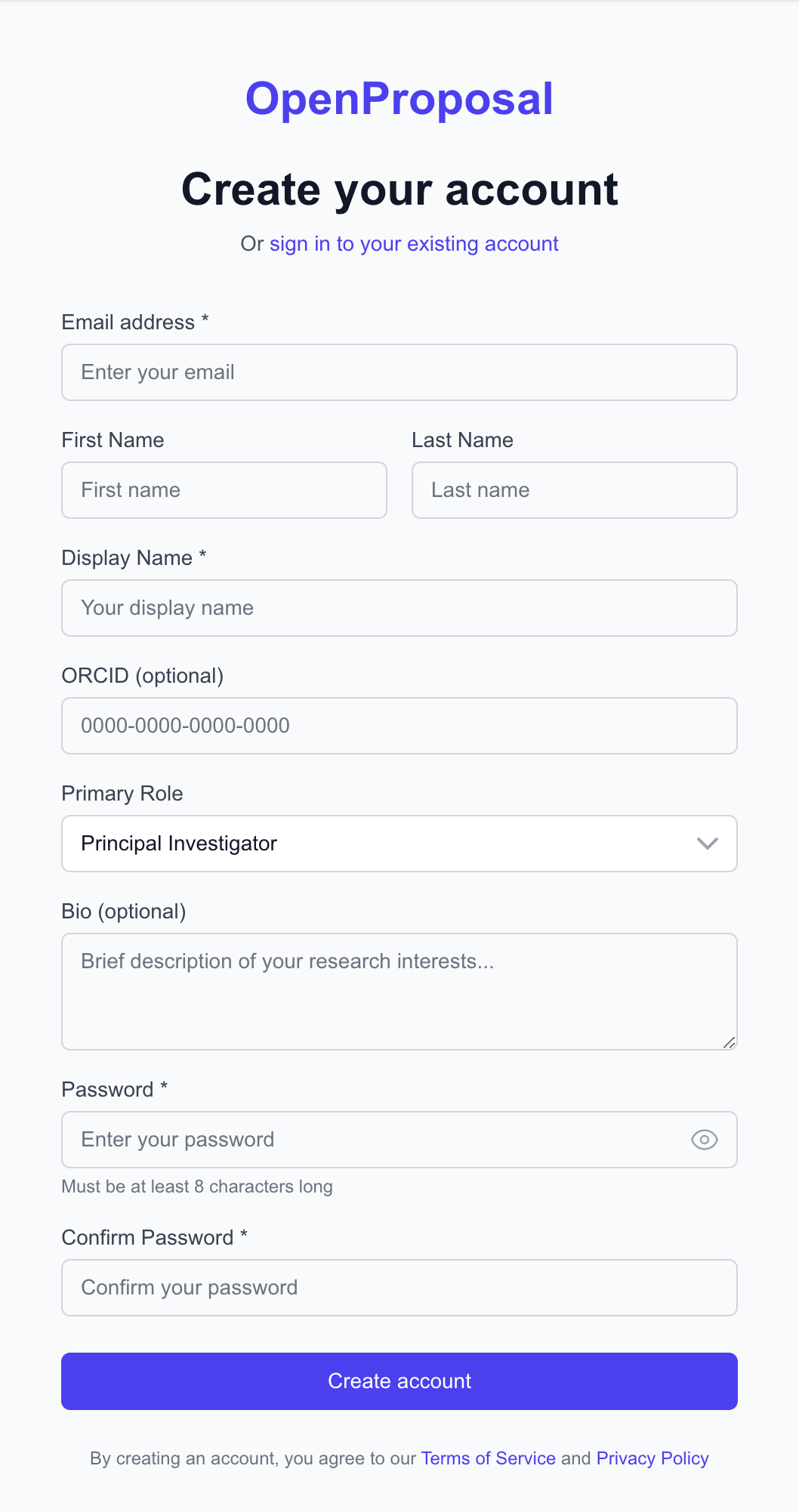}
    \caption{User registration interface with role-based account creation and academic credential integration.}
    \label{fig:signup}
\end{figure}

Getting started with the platform involves a registration process tailored to academic users. We ask for the usual academic credentials - institutional affiliation, research areas, and optionally an ORCID ID for identity verification. The registration flow, shown in Figure~\ref{fig:signup}, breaks this information collection into manageable steps so users don't get overwhelmed.

Different user types (PIs, reviewers, agency staff) see different features once they're logged in, but everyone goes through the same basic registration process. This keeps things simple while ensuring people only access the parts of the system relevant to their role.

\begin{figure*}[htbp]
    \centering
    \includegraphics[width=\linewidth]{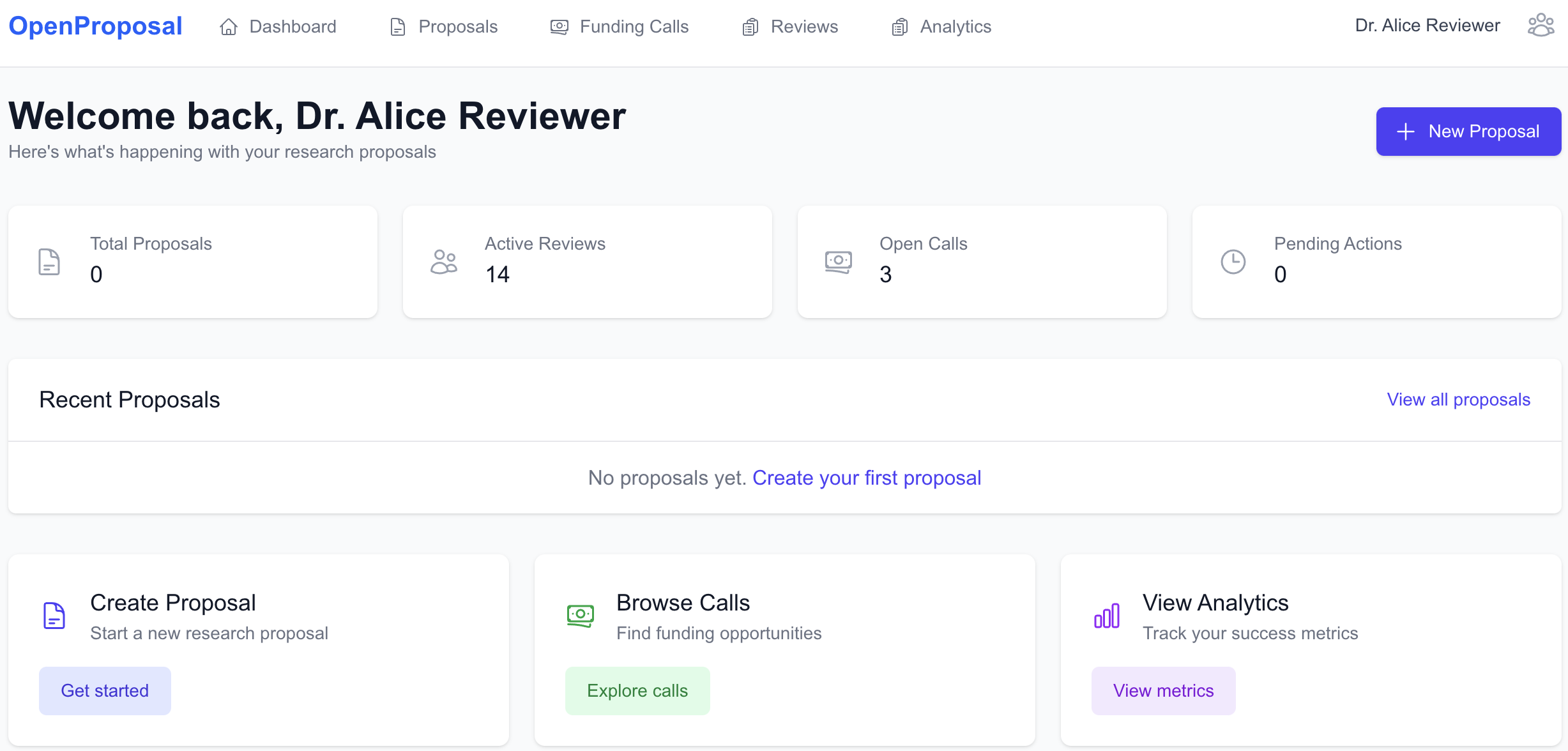}
    \caption{User dashboard showing integrated proposal management, review tracking, and performance analytics.}
    \label{fig:dashboard}
\end{figure*}

\subsubsection{Dashboard and Information Architecture}

The main dashboard (Figure~\ref{fig:dashboard}) serves as home base for all platform activities. We've organized it to show the most important information at a glance - current proposals, pending reviews, upcoming deadlines, and key statistics.

Rather than cramming everything onto one screen, we use charts and quick-access buttons to help users jump to their most common tasks. Whether someone needs to submit a new proposal, accept a review assignment, or check on results, they can get there in just a click or two. The interface works well on both desktop and mobile devices, which is important since many academics work from various locations.

\subsubsection{Funding Opportunity Discovery}
\begin{figure*}[htbp]
    \centering
    \includegraphics[width=\linewidth]{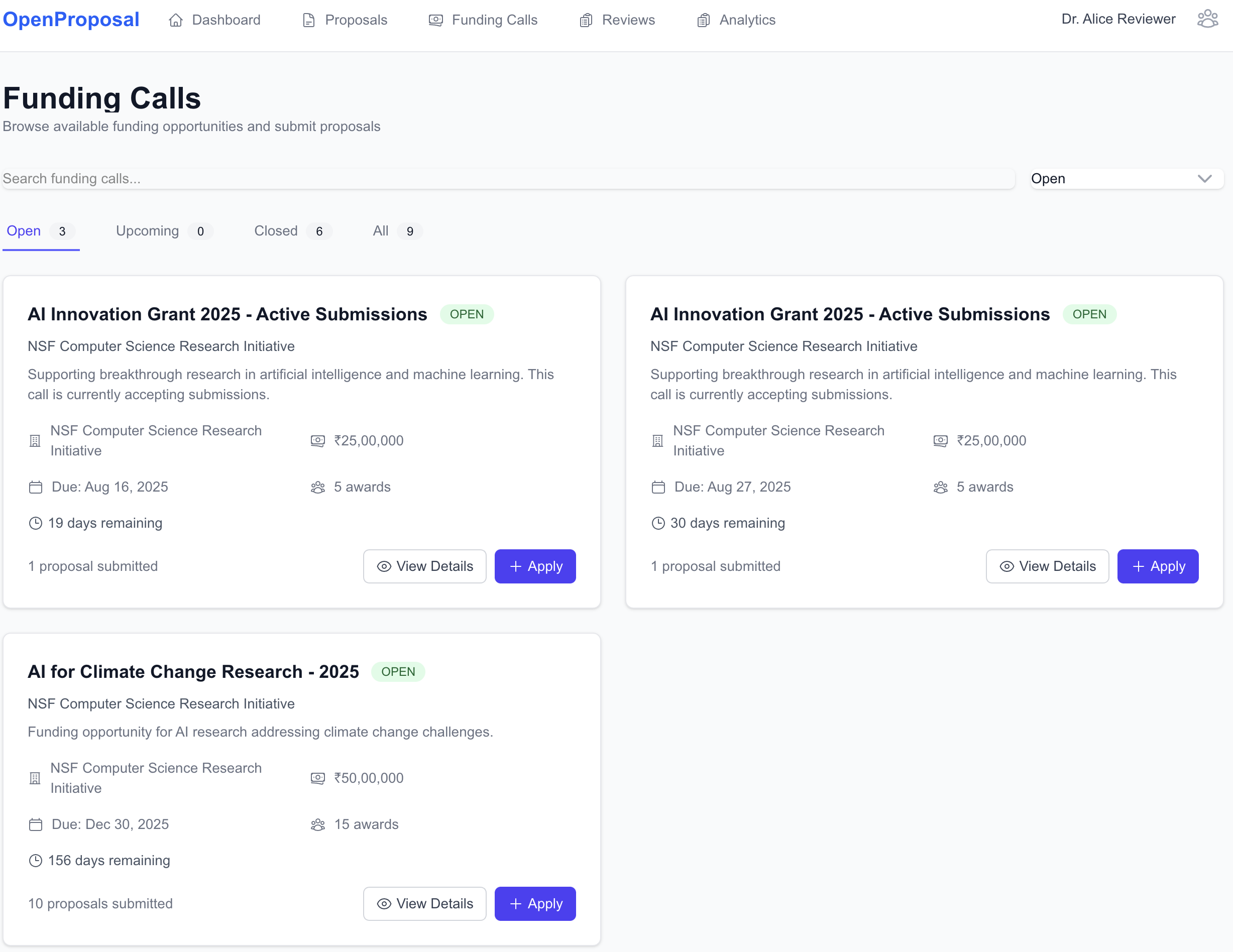}
    \caption{Funding calls discovery interface with advanced filtering and real-time application tracking.}
    \label{fig:funding-calls}
\end{figure*}

Finding relevant funding opportunities can be time-consuming, so we built search and filtering tools to help researchers quickly identify calls that match their work. The funding calls page (Figure~\ref{fig:funding-calls}) shows all the key details - which agency is offering funding, how much money is available, when applications are due, and how many people have already applied.

The filtering system lets users narrow down opportunities by research area, funding amount, application deadline, or agency type. We also show real-time application numbers, which gives researchers a sense of how competitive each call might be. This transparency helps people make more strategic decisions about where to invest their proposal-writing time.

\begin{figure*}[htbp]
    \centering
    \includegraphics[width=0.8\linewidth]{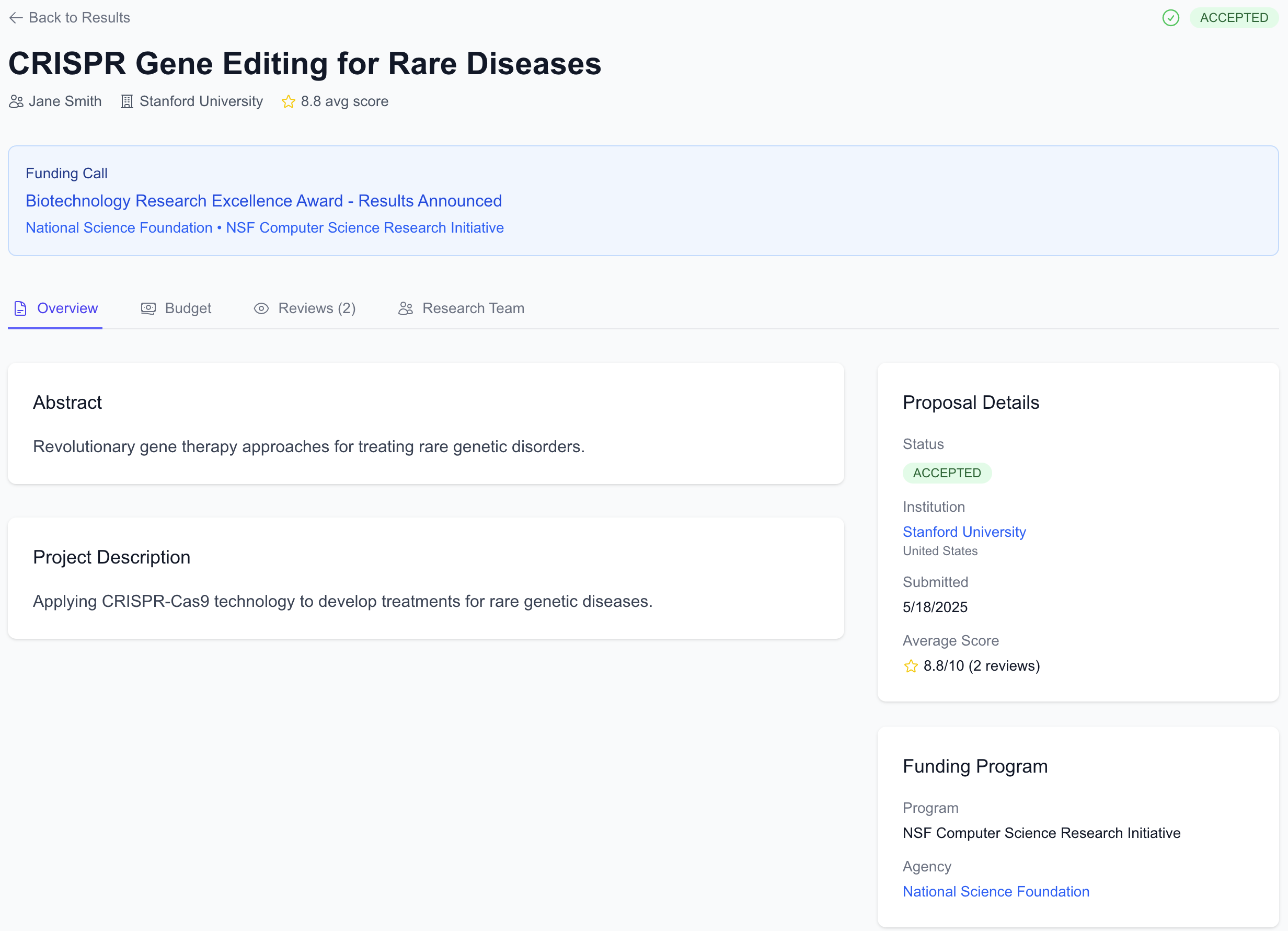}
    \caption{Comprehensive proposal detail interface with organized information architecture and review integration.}
    \label{fig:proposal-detail}
\end{figure*}

\begin{figure}[htbp]
    \centering
    \includegraphics[width=0.9\linewidth]{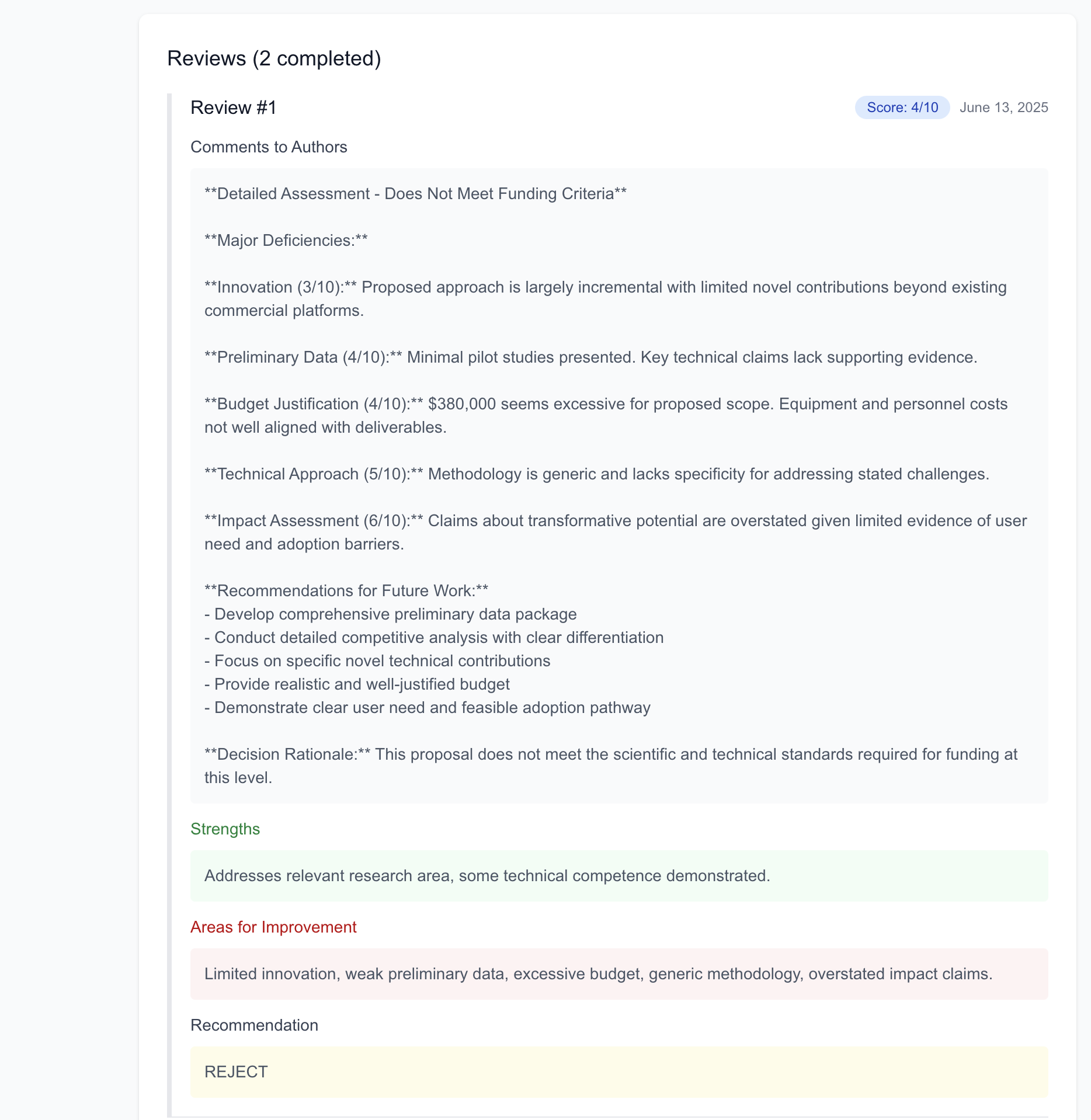}
    \caption{Review interface showing detailed evaluation criteria, scoring mechanisms, and structured feedback collection.}
    \label{fig:reviews}
\end{figure}

\subsubsection{Proposal Presentation and Review Interface}
We've put considerable thought into how proposals are displayed and reviewed. The proposal detail page (Figure~\ref{fig:proposal-detail}) uses tabs to organize different sections - the main research description, budget information (when allowed), team details, and review feedback. This prevents information overload while keeping everything easily accessible.

For reviewers, we've created evaluation forms that balance structure with flexibility (Figure~\ref{fig:reviews}). Reviewers can give numerical scores on different criteria and provide detailed written feedback for each aspect of the proposal. The forms are standardized enough to ensure consistency, but flexible enough to accommodate the nuances of different research fields.

\subsubsection{Transparent Results Publication}
\begin{figure}[h]
    \centering
    \includegraphics[width=\linewidth]{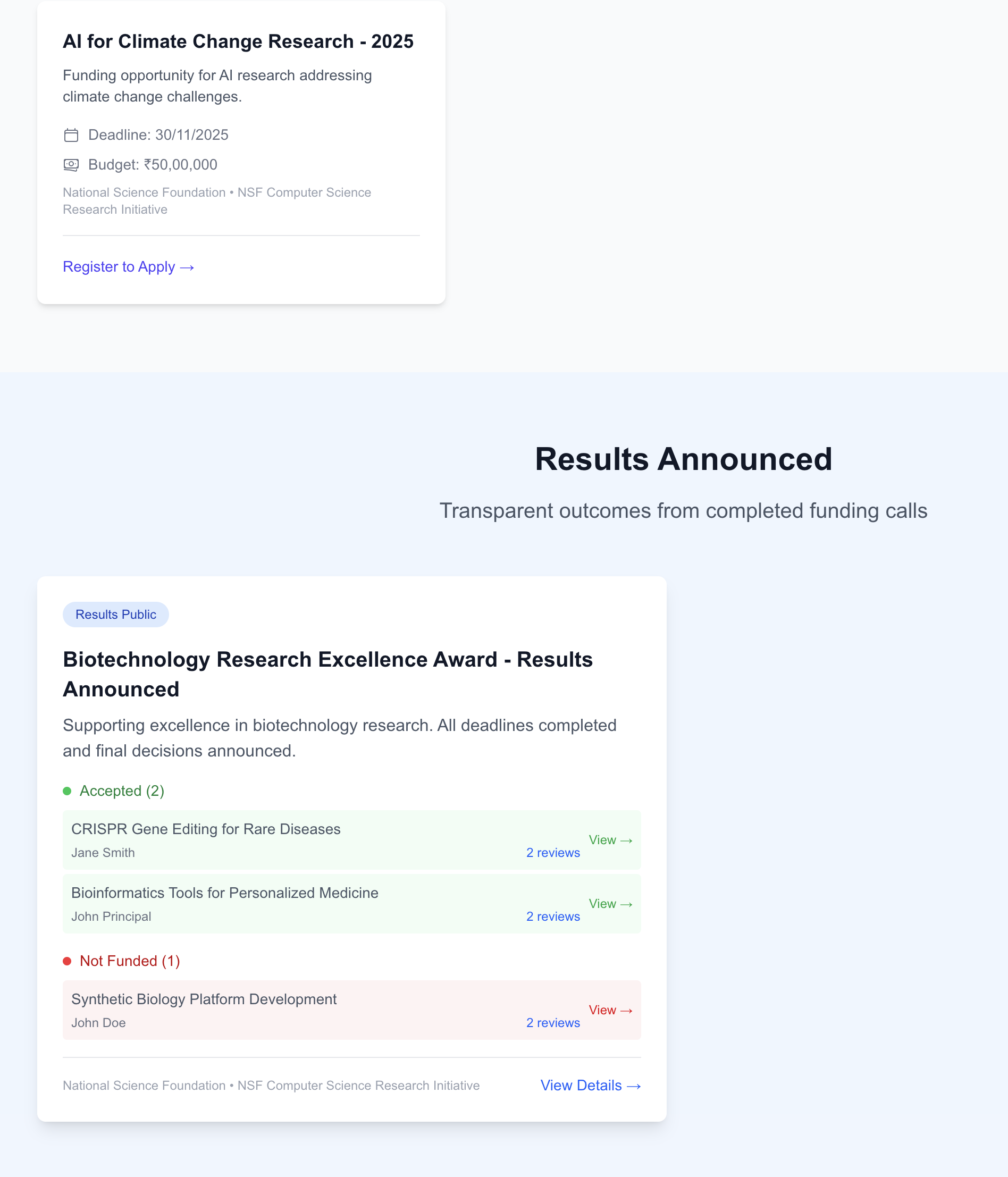}
    \caption{Results publication interface showing transparent funding outcomes with comprehensive review data access.}
    \label{fig:results}
\end{figure}

When funding decisions are finalized, we publish the outcomes in a way that maximizes learning while respecting privacy (Figure~\ref{fig:results}). Both funded and unfunded proposals are listed with their review scores, decision explanations, and links to the full review discussions.

This creates valuable learning opportunities for the research community. Researchers can see what kinds of projects get funded, understand how review criteria are applied in practice, and get ideas for improving their own proposals. The aggregated data also reveals broader patterns in funding - which research areas are thriving, which institutions are most successful, and how review standards evolve over time.

\textbf{Note on Data}: The dataset used in this study was synthetically generated for the purpose of experimentation and evaluation. It is intended solely for illustrative and testing purposes and does not correspond to any real individuals, entities, or events.

\section{Expected Benefits and Impact Analysis}

\textbf{Accountability.} Public reviews in OpenProposal create accountability mechanisms that should improve review quality. When reviewers know their feedback will be visible, they provide more thoughtful and constructive evaluations. This transparency effect has been demonstrated in OpenReview contexts, where public reviews show higher quality than traditional anonymous reviews.

The visibility of the review process also enables community oversight of funding decisions, reducing bias or conflicts of interest in outcomes.

\textbf{Community Engagement and Scientific Rigor.} OpenProposal allows domain experts beyond assigned reviewers to contribute insights and identify issues with proposals. This approach can catch technical errors, highlight relevant prior work, or suggest alternative approaches that assigned reviewers miss.

The platform's discussion mechanisms enable iterative improvement of proposals through community feedback, leading to higher-quality funded research. This approach aligns with open science principles and leverages the distributed expertise of the research community.

\textbf{Author Response Mechanisms.} The integration of rebuttal mechanisms addresses a limitation of traditional funding review: applicants cannot respond to reviewer misunderstandings or errors. OpenProposal enables Principal Investigators to clarify methodological details, correct factual errors, or provide additional evidence that addresses reviewer concerns.

This bidirectional communication is valuable for proposals at funding thresholds, where minor clarifications or additional evidence might influence funding decisions. The rebuttal process ensures that funding decisions are based on complete information rather than reviewer misconceptions.

\textbf{Data-Driven Insights for System Improvement.} OpenProposal's open data approach enables research on peer review effectiveness and bias patterns in funding allocation. By making review data publicly available (with privacy protections), the platform creates opportunities for meta-research on review quality, reviewer expertise matching, and fairness across different demographic groups and research domains.

This data availability supports evidence-based improvements to funding systems, including:

\begin{itemize}
\item Analysis of reviewer expertise effects on review quality and accuracy
\item Longitudinal studies of funding decision accuracy through project outcome tracking
\item Investigation of biases in funding allocation across institutions, demographics, and research areas
\item Evaluation of rebuttal effectiveness in improving decision accuracy
\end{itemize}

These insights can inform policy improvements at funding agencies and contribute to more effective research funding systems.

\section{Limitations}

\textbf{System Limitations and Adoption Challenges.} OpenProposal faces practical limitations that constrain its immediate applicability. The platform represents a proof-of-concept implementation that has not been validated at the scale required for major funding agencies processing thousands of proposals annually. Our technical infrastructure lacks the security auditing, integration capabilities, and performance optimization necessary for production deployment with established funding organizations.

The system faces institutional adoption barriers. Funding agencies operate within established legal frameworks, organizational cultures, and review processes that resist transparency initiatives. The transition from closed to open review requires policy revisions, staff training, and change management that many agencies may be reluctant to undertake. The platform's success depends on achieving adoption, creating a chicken-and-egg problem where agencies await peer adoption before committing resources.

\textbf{Potential Negative Consequences.} Transparent review systems introduce risks that traditional closed systems avoid. Public reviews may reduce participation from experienced researchers who currently serve as reviewers partly due to anonymity protection. Senior reviewers may become reluctant to provide feedback that could damage professional relationships, leading to self-censorship and performative rather than evaluative reviews.

The platform creates privacy risks despite efforts to protect information. In competitive research areas, public proposal details may compromise intellectual property or provide competitor advantages. Even with budget redaction, the combination of public proposals, institutional affiliations, and review discussions may reveal more information than intended.

Public review systems create opportunities for gaming. Research groups with large networks may exploit community engagement features to generate support, disadvantaging researchers from smaller institutions. Public disagreements between reviewers may escalate into professional conflicts, particularly in fields with existing methodological disputes.

\textbf{Evaluation and Validation Gaps.} This work presents a system without empirical validation of its claimed benefits. We cannot demonstrate that transparent funding review actually improves review quality, reduces bias, or enhances fairness without controlled studies comparing transparent and traditional review outcomes. Our benefits remain hypothetical until validated through evaluation with real funding decisions and measurable outcomes.

The platform's emphasis on community engagement and public discussion reflects assumptions about academic cultures that may not generalize across different countries, funding scales, or research domains. Without empirical testing across contexts, the system's broader applicability remains uncertain.

\textbf{Assessment of Contributions.} OpenProposal represents an adaptation of existing transparency principles to funding review rather than a technical or methodological innovation. Our contribution lies in demonstrating the feasibility of such adaptation and identifying design considerations for balancing transparency with funding-specific privacy requirements.

The gap between technical implementation and real-world impact remains substantial. Achieving transparency in funding requires not only technological solutions but also policy changes, cultural shifts, and institutional commitments that extend beyond our current work scope. Future work must include controlled studies with cooperating funding agencies, privacy analysis, and longitudinal evaluation of review quality and fairness outcomes before the platform's benefits can be established.

\section{Conclusion and Future Work}

We have presented OpenProposal, a working platform that demonstrates how transparency principles from academic peer review can be successfully adapted to research funding evaluation. Our system tackles real limitations in current funding processes by implementing public reviews, author rebuttals, and community engagement mechanisms, while thoughtfully protecting sensitive information such as budget details. The platform's technical implementation, built on modern web technologies including Next.js, React, and Prisma, proves that transparent funding review systems are not only possible but practical to build and deploy.

While challenges certainly exist for widespread adoption, including institutional inertia, privacy considerations, and the need for empirical validation these are solvable problems rather than fundamental barriers. We believe transparent funding review will enhance review quality and fairness, building on the demonstrated success of OpenReview in academic publishing. Our platform provides a solid technical foundation for future empirical research on funding review processes, and we're optimistic that validation studies will confirm the benefits we anticipate.

The path forward includes collaboration with progressive funding agencies to conduct pilot studies, systematic evaluation of our privacy protection mechanisms, and controlled experiments measuring review quality and fairness outcomes. Longer-term goals include integration with institutional research management systems and advanced reviewer-proposal matching techniques. OpenProposal represents the beginning of a transformation in funding transparency rather than just another incremental improvement. Our work provides both a practical starting point and the technical infrastructure needed to make transparent funding review a reality for the research community.

\begin{acks}

This work is partially funded by DST Center for Policy Research, NISER. This work has benefited from the use of AI language tools (LLMs) for non-substantive editing, including phrasing and grammatical correction. All intellectual content is original and authored by the contributors.

\end{acks}

\bibliographystyle{ACM-Reference-Format}
\bibliography{main}

\appendix

\end{document}